\documentstyle[11pt,newpasp,twoside,epsf]{article}
\begin{document}

\title {An Overview of the Performance and Scientific Results from the Chandra X-Ray Observatory (CXO)\\ }

\author{M. C. Weisskopf}
\affil{NASA Marshall Space Flight Center, MSFC, AL 35812}
\author{B. Brinkman}
\affil{SRON, Sorbonnelaan 2, 3584 CA Utrecht, the Netherlands}
\author{C. Canizares}
\affil{MIT, 77 Massachusetts Avenue, Cambridge MA 02139-4307}
\author{G. Garmire}
\affil{PSU, 525 Davey Lab, University Park, PA 16802}
\author{S. Murray}
\affil{SAO, 60 Garden Street, Cambridge MA 02138}
\author{L. P. Van Speybroeck} 
\affil{SAO, 60 Garden Street, Cambridge MA 02138}

\begin {abstract}

The Chandra X-Ray Observatory (CXO), the x-ray component of NASA's Great Observatories, was launched on 1999, July 23 by the Space Shuttle Columbia.  
After satellite systems activation, the first x-rays focussed by the telescope were observed on 1999, August 12.  
Beginning with the initial observation it was clear that the telescope had survived the launch environment and was operating as expected. 
Despite an initial surprise due to the discovery that the telescope was far more efficient for concentrating CCD-damaging low-energy protons than had been anticipated, the observatory is performing well and is returning superb scientific data.  
Together with other space observatories, most notably XMM-Newton, it is clear that we have entered a new era of discovery in high-energy astrophysics.
\end {abstract}


\section {Introduction} 

The Chandra X-Ray Observatory (CXO), formerly known as the Advanced X-Ray Astrophysics Facility (AXAF), has joined the Hubble Space Telescope (HST) and the now defunct Compton Gamma-Ray Observatory (CGRO) as one of NASA's "Great Observatories".  
Chandra provides unprecedented capabilities for sub-arcsecond imaging, spectrometric imaging, and for high-resolution dispersive spectroscopy over the band 0.08-10 keV (15-0.12 nm).  
Therefore, a wide variety of high-energy phenomena in an all-encompassing range of astronomical objects is being observed.

Chandra is a NASA facility that provides scientific data to the international astronomical community in response to scientific proposals for its use.  
The Observatory is the product of the efforts of many organizations in the United States and Europe.  
NASA Marshall Space Flight Center (MSFC, Huntsville, Alabama) manages the Project and provides Project Science; TRW Space and Electronics Group (Redondo Beach, California) served as prime contractor; the Smithsonian Astrophysical Observatory (SAO, Cambridge, Massachusetts) provides technical support and is responsible for ground operations including the Chandra X-ray Center (CXC) which distributes and archives the data.  
There are five scientific instruments (\S\ref{ss:xraysys}) aboard the Observatory.  

In 1977, NASA/MSFC, in collaboration with SAO, performed a study which led to the definition of the mission.
This study was a result of an unsolicited proposal submitted to NASA in 1976 by Prof. R. Giacconi and Dr. H. Tananbaum.
Since then, much has transpired, including the highest recommendation by the National Academy of Sciences Astronomy Survey Committee, selection of the instruments, selection of the prime contractor, demonstration of the optics, restructuring of the mission, and ultimately the launch.

We begin by briefly describing the Chandra systems (\S\ref{s:systems}). 
We then describe Chandra's on-orbit performance (\S\ref{s:onorbit}) and highlight certain scientific results (\S\ref{s:scires}).

\section {Chandra Systems} \label{s:systems}

\subsection {Mission and Orbit} \label{ss:mission}

The Space Shuttle {\sl Columbia} launched and deployed the Observatory into a low earth orbit at an altitude of about 240 km.   
Subsequently, an Inertial Upper Stage, a two-stage solid-fuel rocket booster developed by the Boeing Company Defense and Space Group (Seattle, Washington), propelled the Chandra flight system into a highly elliptical transfer orbit.
Subsequently, over a period of days, Chandra's Internal Propulsion System placed the observatory into its initial operational orbit - 140,000-km apogee and 10,000-km perigee, with a 28.5$^o$ inclination.
The highly elliptical orbit, with a period of 63.5 hours, yields a high observing efficiency.
The fraction of the sky occulted by the earth is small, as is the fraction of the time when the detector backgrounds are high as the Observatory dips into Earth's radiation belts.
Consequently, more than 70\% of the time is useful and uninterrupted observations lasting more than 2 days are possible.

The specified design life of the mission is 5 years; however, the only expendable (gas for maneuvering) is sized to allow operation for much more than 10 years and the orbit will remain useful for decades.

\begin {figure} 
\caption {\label{fig:flightsys} 
Expanded view of the Chandra flight system, showing several subsystems.  TRW drawing.}
\end {figure}

\subsection {Spacecraft}

The spacecraft is made up of:
\begin{itemize}
\item [1.]  The Pointing Control and Aspect Determination subsystem which performs on-board attitude determination, solar-array control, slewing, pointing and dithering control, and momentum management.
\item [2.]  The Communication, Command, and Data Management subsystem which performs communications, command storage and processing, data acquisition and storage, and computation support, timing reference, and switching of primary electrical power for other systems or subsystems.
\item [3.]  The Electrical Power Subsystem  which generates, regulates, stores, distributes, conditions, and controls the primary electrical power.
\item [4.]  The Thermal Control Subsystem which furnishes passive thermal control (where possible), heaters, and thermostats.
\item [5.]  The Structures and Mechanical Subsystem which encompasses the spacecraft structures, mechanical interfaces among the spacecraft subsystems and with the telescope system and external structures.
\item [6.]  The Propulsion Subsystem which comprises the Integral Propulsion Subsystem - deliberately disabled once final orbit was obtained - and the Momentum Unloading Propulsion Subsystem.
\item [7.]  The flight software which implements algorithms for attitude determination and control, command and telemetry processing and storage, and thermal and electrical power monitoring and control.
\end{itemize}

\subsection {Telescope system} \label{sss:hrmasys}

Kodak integrated the Telescope System. 
Its principal elements are the High-Resolution Mirror Assembly (HRMA, \S\ref{sss:hrma}) and the Optical Bench Assembly (OBA).
Composite Optics Incorporated (COI, San Diego, California) developed the critical light-weight composite materials for  the optical bench and for other Chandra structures.
The Telescope System also provides mounts and mechanisms for the Chandra Observatory's 2 objective transmission gratings (\S\ref{sss:gratings}).
In addition, Ball Aerospace and Technologies Corporation (Boulder, Colorado) fabricated the Aspect Camera Assembly (Michaels 1998), a visible-light telescope and CCD camera which attaches to, and is coupled with, the HRMA through a fiducial-light transfer system, which maps the x-ray focal plane onto the sky.

\subsection {Integrated Science Instrument Module} \label{ss:sisys}

Ball Aerospace and Technologies Corporation also built the Science Instrument Module (SIM) - Skinner \& Jordan (1997) - which includes mechanisms for focussing  and translating Chandra's focal-plane science instruments (\S\ref{sss:fpsi}).
The translation is necessary as the instruments cannot realistically share the focal plane and must be translated into position at the telescope focus.

\subsection {Electron Proton Helium Instrument (EPHIN)} 

Mounted on the spacecraft and near the HRMA is a particle detector: the Electron, Proton, Helium INstrument (EPHIN).
The EPHIN instrument was built by the Institut f\"ur Experimentelle und 
Angewandte Physik, University of Kiel, Germany, and a forerunner 
was flown on the SOHO satellite. 

EPHIN consists of an array of 6 silicon detectors with anti-coincidence. 
The instrument is sensitive to electrons in the energy range 250 keV - 10 
MeV, and hydrogen and helium isotopes in the energy range 5 - 53 MeV/nucleon. 
Electrons above 10 MeV and nuclei above 53 MeV/nucleon are registered with reduced 
capability to separate species and to resolve energies. The field of view is 83 degrees 
with a geometric factor of 5.1 cm$^2$ sr. A detailed instrument description is given in Mueller-Mellin et al. (1995).

EPHIN is used to monitor the local charged particle environment as part of the scheme to protect the focal-plane instruments from particle radiation damage. Clearly EPHIN is also a scientific experiment in its own right.

\subsection {X-Ray Subsystems} \label{ss:xraysys}

Chandra's x-ray subsystems are the High-Resolution Mirror Assembly (HRMA), the objective transmission gratings, and the focal-plane science instruments.

\subsubsection {High-Resolution Mirror Assembly (HRMA)} \label{sss:hrma}

Hughes Danbury Optical Systems (HDOS, Danbury, Connecticut) precision figured and superpolished the 4-mirror-pair grazing-incidence x-ray optics out of Zerodur blanks from Schott Glaswerke (Mainz, Germany).
Optical Coating Laboratory Incorporated (OCLI, Santa Rosa, California) coated the optics with iridium, chosen for high x-ray reflectivity and chemical stability.
The Eastman Kodak Company (Rochester, New York) aligned and assembled the mirrors into the 10-m focal length High-Resolution Mirror Assembly (HRMA, Figure~\ref{fig:hrma}).  
The forward contamination cover houses 16 radioactive sources used for verifying transfer of the flux scale from the ground to orbit (Elsner et al. 1994, 1998, 2000).

\begin {figure} 
\caption{Photograph of the High-Resolution Mirror Assembly during alignment and assembly at Kodak.
In the photo, 7 of the 8 mirrors are already attached to the central aperture plate.
Photograph is from Kodak.\label{fig:hrma}}
\end {figure}

\subsubsection {Objective transmission gratings} \label{sss:gratings}

Aft of the HRMA are 2 objective transmission gratings (OTGs) - the Low-Energy Transmission Grating (LETG) and the High-Energy Transmission Grating (HETG).
Positioning mechanisms are used to insert either OTG into the converging beam where they disperse the x-radiation onto the focal plane. Figure~\ref{fig:gratings} shows the gratings mounted behind the HRMA in their retracted position.

\begin {figure} 
\caption {\label{fig:gratings} 
Photograph of the LETG and HETG mounted to the spacecraft structure.  Photograph is from TRW.}
\end {figure}

\paragraph{Low-Energy Transmission Grating (LETG)}

The Space Research Institute of the Netherlands (SRON, Utrecht, Netherlands) and the Max-Planck-Instit\"ut f\"ur extraterrestrische Physik (MPE, Garching, Germany) designed and fabricated the LETG.
The 540 grating facets, mounted 3 per module, lie tangent to the Rowland toroid  which includes the focal plane.
With free-standing gold bars of about 991-nm period, the LETG provides high-resolution spectroscopy from 0.08 to 2 keV (15 to 0.6 nm).

\paragraph{High-Energy Transmission Grating (HETG)}

The Massachusetts Institute of Technology (MIT, Cambridge, Massachusetts) designed and fabricated the HETG.
The HETG employs 2 types of grating facets~--- the Medium-Energy Gratings (MEG), mounted behind the HRMA's 2 outermost shells, and the High-Energy Gratings (HEG), mounted behind the HRMA's 2 innermost shells~--- oriented at  different dispersion directions.
With polyimide-supported gold bars of 400-nm and 200-nm periods, respectively, the HETG provides high-resolution spectroscopy from 0.4 to 4 keV (MEG, 3 to 0.3 nm) and from 0.8 to 8 keV (HEG, 1.5 to 0.15 nm).

\subsubsection {Focal-plane science instruments} \label{sss:fpsi}

The Integrated SIM (\S \ref{ss:sisys}) houses Chandra's 2 focal-plane science instruments~--- the (microchannel-plate) High-Resolution Camera (HRC) and the (Charged Coupled Device - CCD) Advanced CCD Imaging Spectrometer (ACIS).
Each instrument provides both a so called (as all the detectors are imagers) imaging detector (I) and a spectroscopy detector (S), the latter designed especially to serve as a readout for the photons dispersed by the transmission gratings. 

\paragraph{High-Resolution Camera (HRC)}

SAO designed and fabricated the HRC (Murray et al. 2000) shown in Figure~\ref{fig:hrc}.
Made of a single 10-cm-square microchannel plate, the HRC-I provides high-resolution imaging over a 31-arcmin-square field of view.
Comprising 3 rectangular segments (3-cm $\times$ 10-cm each) mounted end-to-end along the OTG dispersion direction, the HRC-S serves as the primary read-out detector for the LETG.
Both detectors are coated with a cesium--iodide photocathode and covered with aluminized-polyimide UV/ion shields.

\begin {figure} 
\caption {\label{fig:hrc} 
Photograph of the HRC.
The HRC-I (imager) is at the bottom; the HRC-S (the readout for the LETG), at the top.
}
\end {figure}

\paragraph {Advanced CCD Imaging Spectrometer (ACIS)} 
The Pennsylvania State University (PSU, University Park, Pennsylvania) and MIT designed and fabricated the ACIS (Figure~\ref{fig:acis}) with CCDs produced by MIT Lincoln Laboratory (Lexington, Massachusetts). Some subsystems and systems integration was provided by Lockheed--Martin Astronautics (Littleton, Colorado).
Made of a 2-by-2 array of large-format, front-illuminated (FI), 2.5-cm-square, CCDs, ACIS-I provides high-resolution spectrometric imaging over a 17-arcmin-square field of view.
ACIS-S, a 6-by-1 array of 4 FI CCDs and two back-illuminated (BI) CCDs mounted along the OTG dispersion direction, serves both as the primary read-out detector for the HETG, and, using the one BI CCD which can be placed at the aimpoint of the telescope, also provides high-resolution spectrometric imaging extending to lower energies but over a smaller (8-arcmin-square) field than ACIS-I.
Both ACIS detectors are covered with aluminized-polyimide optical blocking filters.

\begin {figure} 
\caption {\label{fig:acis} 
Photograph of the focal plane of ACIS, prior to installation of the optical blocking filters.
The ACIS-I  is at the bottom; the ACIS-S (the readout for the HETG), at the top.
}
\end {figure}

\section {On-orbit Performance} \label{s:onorbit}

Chandra's mission is to provide high-quality x-ray data. 
Chandra's performance advantage over other x-ray observatories is analogous to that of the HST over ground-based telescopes.
The effective area of the Chandra mirror is shown in
Figure~\ref{fig:sys-eff-area}; it is approximately 800~${\rm cm}^2$ at energies below 2~keV, and approximately 400~${\rm cm}^2$ between 2 and 5 keV. 
Figure~\ref{fig:sys-eff-area} also shows
the effective area convolved with ACIS quantum efficiencies. 

\begin {figure} [htb]
\caption {\label{fig:sys-eff-area} 
Chandra on-axis mirror and mirror/ACIS effective areas.}
\end {figure}

\subsection {Imaging} \label{sss:imaging}

The angular resolution of Chandra is significantly better than any previous, current, or even currently-planned x-ray observatory.
Figure~\ref{fig:casa} qualitatively, yet dramatically, illustrates this point by comparing the early Chandra image of the supernova remnant Cassiopeia-A, based on about 2700 s of data, with a $\sim$200,000 s ROSAT image.
(Prior to Chandra, the ROSAT observatory represented the state of the art in high-resolution x-ray imaging.)
The improvement is dramatic, and the point source at the center~--- undetected in the ROSAT image~--- simply leaps out of the Chandra image.
\vspace{0.10in}

\begin {figure} 
\caption {\label{fig:casa} 
Chandra (left) and ROSAT (right) images of CAS-A.}
\end {figure}

Quantitatively, Chandra's point spread function (PSF), as measured during ground calibration, had a full width at half-maximum less than 0.5 arcsec and a half-power diameter less than 1 arcsec.
The prediction for the on-orbit encircled-energy fraction was that a 1-arcsec-diameter circle would enclose at least half the flux from a point source.
A relatively mild dependence on energy resulting from diffractive scattering by surface microroughness attested to the excellent superpolished finish.
The ground measurements were taken under environmental conditions quite different than those encountered on-orbit.
The effects of gravity and the finite distance and size of the various x-ray sources were unique to the ground calibration.
On the other hand, on the ground there was no observatory motion to deal with.
On-orbit, the performance folds in the spatial resolution of the flight detectors and any uncertainties in the aspect solution which determines, post-facto, the direction the observatory was pointing relative to the instruments and to celestial coordinates.

The HRC has the best spatial resolution ($\sim$ 20$\mu$m, $\sim$0.4 arcsec) of the two imaging instruments and thus is best matched to the telescope.
Figure~\ref{fig:eehrci} illustrates the extrapolation of the ground calibration to on-orbit and compares predictions at two energies with an observed PSF. More details as to the on-orbit imaging performance may be found in Jerius et al. (2000).
The performance of the aspect camera and the attitude control system is discussed by Aldcroft et al. (2000) and Cameron et al. (2000). 

\begin {figure} 
\caption {\label{fig:eehrci} 
The predicted and observed encircled energy as a function of radius for an on-axis point source as observed with the HRC-I.
The calculations, performed at two energies (0.277 keV and 6.40 keV) include a realistic (0.22") estimate of the contribution from the aspect solution.
Flight data are from the calibration observation of AR Lac.
Figure produced by Chandra Telescope Science.}
\end {figure}

Finally, it is interesting that the use of the zeroth order image for the observations of extremely bright sources, which would otherwise saturate the detectors and/or the telemetry, has proven quite useful.
The utility for such observations is illustrated in Figure~\ref{fig:3c273} where we show an image with the LETG inserted. 
Both the jet and the central source of 3C273 are clearly resolved.

\begin {figure} 
\caption {\label{fig:3c273} 
Image of the dispersed spectrum, including zeroth order, of 3C273.
The jet is clearly resolved in the lower right hand portion of the figure.
The six spikes emanating from the central image are due to dispersion by the facet holders.
Image courtesy Jeremey Drake and LETGS team.}
\end {figure}

\subsection {HRC} \label{sss:hrcperf}

The HRC is performing close to what was expected pre-launch, despite a few anomalies discussed below.

\subsubsection{Background}

The HRC-I on-orbit counting rate is about 250 c/s ($\sim 2$ cts/s/cm$^{2}$). These are mostly cosmic ray events and are detected in the anti-coincidence shield (AC). 
The on board veto function has been activated and reduces the valid event rate to about 50 c/s. 

In the case of the HRC-S, enabling the anti-coincidence shield does
not reduce the background as expected. 
The problem appears to be a timing error in the electronics that can not be
changed from the ground. 
The HRC-S trigger pulses arrive earlier than the AC signals and are not held long enough to trigger a coincidence, resulting in telemetry saturation.
To cope, a ``spectroscopic region'' has been defined which is a strip of the detector, 9.6 mm wide, centered on the nominal spectroscopy aim point. 
This region is about 1/2 of the total HRC-S area and reduces the valid count rate to about 120 c/s, well below the telemetry limit.

An (on-ground) event screening algorithm has been developed to help in identifying non-X-ray events and further reduces the background for data from both HRC-I and -S. 
For HRC-I, e.g., this results in a background rate of about 0.8 cts/arcsec$^{2}$ for a $10^{5}$ second integration.
For HRC-S, this approach reduces the background to $\sim 13$ counts in a 0.06\AA\ spectral ``slice'' per $10^{5}$ second integration. 

\subsubsection{Image Quality}

Even with a reduction in high voltage and MCP gain, there is a
degradation of image quality in HRC-I for the higher amplitude events
due to added electronic noise during event processing. Fortunately
there is enough information in the telemetry stream to correct for this effect. A revised ground system event processing algorithm has been developed
which identifies distorted events and flags them as such. 
The added electronic noise, a systematic effect, can be partially
compensated for and removed. 
Algorithms for making these corrections have been added to the data processing system. 
Using the plate focus data as a test case, the measured encircled energy near the focus was improved from about 40\% within 1 arcsecond diameter to $>60$\%.

A second image artefact is the appearance of ``ghost'' images near
strong sources. 
These ghost images are due to events where the readout amplifiers
are saturated leading to fairly large miscalculation of the event
position. 
The number of such misplaced events is 1 percent or less of the properly
placed events. 
Screening the data for events where the amplifiers are all near their peak value effectively eliminates these ``ghosts''. 

Because the HRC-S gain is lower than for HRC-I, electronic saturation effects are less significant. 
As a result the imaging performance appears to be excellent once the corrections described above for HRC-I have been applied. 

\subsubsection{Efficiency}

The count rates from the sources observed are about what was expected. 
For LMC X-1, Ar Lac and Cas-A the HRC rates are close to the pre-launch predictions. 
There have been no measurable changes to the HRC efficiency since launch.

\subsubsection {HRC Timing} \label{sss:hrctiming}

The HRC was mis-wired so that the time of the event associated with the j-th trigger is that of the previous (j-th -1) trigger. 
If the data from all triggers were routinely telemetered, the mis-wiring would not be problematical and could be dealt with by simply reassigning the time tag  which is nominally accurate to 16 $\mu$sec. 
Since the problem has been discovered, new operating modes have been defined which allow one to telemeter all data whenever the total counting rate is moderate to low, albeit at the price of higher background. 
For very bright sources the counting rate is so high that information associated with certain triggers are never telemetered. 
In this case, the principal reason for dropping events is that the on-board, first-in-first-out (FIFO) buffer fills as the source is introducing events at a rate faster than the telemetry readout. 
Events are dropped until readout commences freeing one or more slots in the FIFO. 
This situation can also be dealt with (Tennant et al. 2001a) and time resolution of the order of a millisecond can be achieved even under these conditions.

More details of the HRC and its performance may be found in Murray et al. 2000,  Kenter et al. (2000), and Kraft et al. (2000).

\subsection{ACIS} \label{sss:acisperf}

As with the HRC, the ACIS instrument is performing well and contributing to the success of the Chandra Mission.

\subsubsection{Background}
The background experienced by the ACIS CCDs shows occasional flares depending upon the orientation of the Observatory's orbit with respect to the Earth's magnetosphere.
The frequency and magnitude of the flares are much more pronounced in the BI CCDs than in the FI CCDs, consistent with the suggestion that the flaring events are caused by low energy protons that enter the observatory through reflection off the very smooth iridium coated mirrors that comprise the HRMA. 
The gate structure of the FI CCDs absorbs most of the protons for the majority of the flares. 

The region of the Hubble Deep Field-North was observed using the ACIS-I array, for 970 ksec in a series of twelve pointings spanning a period of 15 months and thus provides, after removing sources, an excellent representation of the background and its spectrum.
The background (0.5-10.0 keV) was observed to be constant to within about 10\% for eight of the pointings.
The first three pointings were made with the focal plane temperature at -110C and the background was about 10 - 14\% higher than for the background one year later taken at -120C.
A reduction in background was expected at the lower temperature. 
One pointing, ObsId 2344, experienced a highly variable background with a flare increasing the background by a factor of between two and three for about 20 ksec, depending upon the energy band.

The spectrum is shown in Figure~\ref{fig:mergedspectrum}. 
The prominent lines in the background spectrum are from cosmic ray induced fluorescence of the gold-coated collimator, the nickel-coated substrate of the collimator, the silicon in the CCDs and from aluminum used in various places in the housing and the filter coating.
In general the background produced by the fluorescent lines is only about 2.6\% of the background not found in the lines in the soft (0.5-2.0 keV) band and 13.5\% of the flux in the hard (2.0-10.0 keV) band.
Brandt et al. (2001a) have noted that by selecting certain ACIS event grades  it is possible to suppress the background by another 36\% in the soft band (0.5 - 2.0 keV) and by 28\% in the hard band (2.0 - 8.0 keV), while only reducing the source counts by 12\% and 14\% respectively. 

\begin {figure}
\caption {\label{fig:mergedspectrum}  
The spectrum of all x-ray events detected during a 970 ksec exposure to the Chandra Deep Field-North region.}
\end {figure}

\subsubsection {Proton Damage to the Front-Illuminated CCDs} \label{sss:damage}

The ACIS FI CCDs originally approached the theoretical limit for energy resolution at almost all energies, while the BI were of somewhat lesser quality in this regard. 
Subsequent to launch and orbital activation, the energy resolution of the FI CCDs has become a function of the row number, being nearer pre-launch values close to the frame store region and progressively degraded towards the farthest row (Figure~\ref{fig:rows}).
The points are for the FI data and the curves for the BI data.
These data were taken at -120$^o$C.
Note that these curves are representative of the variation with row number, but  do not account for an added row-dependent gain variation which increases the energy resolution by an additional 15-20\% for the larger row numbers.

\begin {figure}
\caption {\label{fig:rows}  
The energy resolution of two of the CCDs (S3 a BI CCD and I3 a FI CCD) as a function of row number.
}
\end {figure}

For a number of reasons, we believe that the damage was caused by low energy protons, encountered during radiation belt passages and reflecting off the x-ray telescope onto the focal plane.
Subsequent to the discovery of the degradation, operational procedures were changed so that the ACIS instrument is not left at the focal position during radiation belt passages.
Since this procedure was initiated, no further degradation in performance has been encountered.
The BI CCDs were not impacted, consistent with the proton-damage scenario, as it is far more difficult for low energy protons to deposit their energy in the buried channels (where damage is most detrimental to performance) of the BI devices, as these channels are near the gates and the gates face in the direction opposite to the HRMA.
The energy resolution for the two BI CCDs remains at their prelaunch values.

\subsection {Grating performance} \label{sss:spec}

The Chandra OTGs allow measurements with spectral resolving power (Figure~\ref{fig:gres}) of $(\lambda/\Delta\lambda = (E/\Delta~E) > 500$ for wavelengths $\lambda > 0.4$ nm (energies $<$ 3 keV). The on-orbit spectral resolution and efficiencies of both the LETG and the HETG were as expected based on pre-launch calibrations. 

\begin {figure} 
\caption {\label{fig:gres}  
Spectral resolving power of the Chandra OTGs.
On-orbit results indicate slightly better performance.
}
\end {figure}

\section {Scientific Results} \label{s:scires}

X-rays result from highly energetic processes - thermal processes in plasmas with temperatures of millions of degrees or nonthermal processes, such as synchrotron emission or scattering from very hot or relativistic electrons.
Consequently, x-ray sources are frequently exotic:

\begin{itemize}
\item Supernova explosions and remnants, where the explosion shocks the ambient interstellar medium or a pulsar powers the emission.
\item Accretion disks or jets around stellar-mass neutron stars or black holes.
\item Accretion disks or jets around massive black holes in galactic nuclei.
\item Hot gas in galaxies and in clusters of galaxies, which traces the gravitational field for determining the mass.
\end{itemize}

Here we give several examples of observations with Chandra which demonstrate  the capability for investigating these processes and astronomical objects through high-resolution imaging (\S~\ref{sss:sciimaging}) and high-resolution spectroscopy (\S~\ref{sss:scispec}).

\subsection {Imaging} \label{sss:sciimaging}

Chandra's capability for high-resolution imaging (\S~\ref{sss:imaging}) enables detailed high-resolution studies of the structure of extended x-ray sources, including supernova remnants (Figure~\ref{fig:casa}), astrophysical jets (Figure~\ref{fig:3c273}), and hot gas in galaxies and clusters of galaxies.
The capability for spectrometric imaging allows studies of structure, not only in x-ray intensity, but in temperature and in chemical composition.
Through observations with Chandra , one has begun to address several of the most exciting topics in contemporary astrophysics. 

\subsubsection{Normal Stars} 
The aspect solution of the spacecraft typically provides absolute sky coordinates to about 1 arc second. 
It is usual for observations exceeding 20 ks to detect several stars in X-rays that have accurate positions (typically 0.4 arcseconds for USNO A2 stars and 0.15 arcseconds for Tyco II Catalog stars). 
Using the stellar positions as a local reference, the x-ray sources can be positioned to about 0.2 to 0.4 arcseconds. 
One example of using such positions is the determination of the positions of the 1000 x-ray sources in the Orion Trapezium region (Figure~\ref{fig:orion}) where the stellar and x-ray positions differ by about 0.3 arcseconds rms (Feigelson et al. 2001). 

\begin {figure}
\caption {\label{fig:orion} 
The ACIS-I image of the Orion Trapezium. The full field is about 16 arcminutes on a side, with the Trapezium stars in the center.
}
\end {figure}

\subsubsection{The Galactic Center}

Precise positioning with Chandra was critical for the unique identification with SgrA* (Baganoff et al. 2001a) in the extremely crowded region of the Galactic Center (Figure~\ref{fig:sgrA}).
The mass of the black hole at SgrA* has been determined from stellar motion to be $2.6 \times 10^{6}$ solar masses (Genzel et al. 2000).      
The X-ray source associated with SgrA* is very faint compared to other galactic nuclei, emitting only about $2\times 10^{33}$ ergs/s. 
A bright flare was detected from this source on 27 October 2000, where the flux increased by over an order of magnitude for about 10 ksec and then rapidly dipped on a time scale of 600 seconds (Baganoff et al. 2001b).

\begin {figure} 
\caption {\label{fig:sgrA} 
The ACIS-I image of the Galactic Center. }
\end {figure}

\subsubsection{Supernova Remnants}
Another example of Chandra's ability to provide high-contrast images of features of low surface brightness is exemplified by the now classic image (Figure~\ref{fig:crab}) of the Crab Nebula and its Pulsar (Weisskopf et al. 2000, Tennant et al. 2001a) which shows the intricate structure produced by the pulsar wind and the synchrotron torus. 

\begin {figure} 
\caption {\label{fig:crab} 
LETGS image of the Crab pulsar and nebula. The nearly horizontal line in the figure is the cross-dispersed spectrum produced by the LETG fine support bars. The nearly vertical line is the dispersed spectrum from the pulsar. 
}
\end {figure}

In cases where there is sufficient signal, the precisely measured point response function of the HRMA permits image deconvolution 
A good example is given by the observation of the recent supernova remnant 1987A in the Large Magellanic Cloud. 
The image is shown in Figure~\ref{fig:snr1987a} where a Lucy-Richardson algorithm has been used to deconvolve the telescope PSF (Burrows et al. 2000). 

\begin {figure} 
\caption {\label{fig:snr1987a} 
The ACIS-S3 image of the supernova remnant SN1987A in the Large Magellanic Cloud. 
The white overlay lines are from a HST image.
}
\end {figure}

\subsubsection{Globular Clusters}
One of the most striking examples of the power of high resolution x-ray  
imaging, is in the spectacular Chandra images of globular clusters.
Figure~\ref{fig:47tuc} is the moderately deep exposure (70 ksec) ACIS-I image 
recently published by Grindlay et al. (2001a) of 47Tuc. The upper panel of  
this "true" color x-ray image (composed of red/green/blue images derived  
from counts recorded  in soft (0.5-1.2 keV), medium(1.2-2 keV), hard (2-6 keV)  
bands) shows the central 2.5' x 2' of the cluster, or approximately  
central 3 core radii.
The enlargement at the bottom of the figure is $\sim30$ arcsec square, and thus the central $\sim0.7$ core radius portion of the cluster.
Some 108 sources are detected in the field excluding the central core, with
L$_{x}>10^{30}$ erg/s. 
Another $>100$ sources are likely present in the central core.

\begin {figure} 
\caption {\label{fig:47tuc} 
Chandra image using ACIS-I3 of 47Tuc. 
The upper image covers the central 2' x 2.5'. 
The enlarged central region is 35" x 35". 
Source identifications shown are MSPs (circles), quiescent LMXBs (circles), CV candidates (squares) and possibly flaring BY Dra systems, or M-S binaries (triangles).
Figure courtesy of Josh Grindlay.
}
\end {figure}

The image, associated spectra, and measured time variability reveals more about the binary content and stellar, as well as dynamical, evolution of a globular cluster than achieved with all previous x-ray observations of globulars combined (and, even arguably, many HST observations as well).
All 16 of the millisecond pulsars (MSPs) recently located from precise pulse timing (Freire et al. 2001), are detected (circles in Figure~\ref{fig:47tuc}).
Their x-ray spectral properties (colors) vs. radio pulsation spindown measures shows them to obey a significantly different L$_{x}$ vs. \.{E} relation than for MSPs in the field (Grindlay et al. 2001b) and 50-100 MSPs are likely detected in the cluster.

The second most abundant x-ray source population in 47Tuc are the long-sought 
accreting white dwarfs, or cataclysmic variables (CVs), marked as squares  
in Figure~\ref{fig:47tuc} for those already identified in deep HST images. 
Many others (primarily blue or whitish color) candidates are present, so that perhaps a third of the Chandra sources are CVs.
A third significant population of x-ray binaries were discovered (and unanticipated) in this image: main sequence star binaries (detached), or so-called BY Draconis stars (the brightest few of which, detected as flaring sources, are marked with triangles).

The observations of other globular clusters with Chandra are beginning to  
be published with a recent example being the nearby core-collapsed cluster  
NGC 6397 (Grindlay et al. 2001c).
This cluster shows a dramatic contrast with 47Tuc: although almost as  
abundant in CVs, it is nearly devoid of MSPs. 
Grindlay et al. (2001c) note that this suggests fundamental differences in the  
relative neutron star vs. white dwarf content, as well as compact binary  
formation history.
Clearly the high resolution x-ray view made possible with Chandra is  
opening a new era in understanding these oldest, and dynamically most  
interesting, stellar systems.

\subsubsection{Normal Galaxies}

In addition to mapping the structure of extended sources and the diffuse emission in galaxies, the high angular resolution permits studies of ensembles of discrete sources, which would otherwise be impossible owing to source confusion.
A beautiful example comes from the observations of the center of M31 (Figure~\ref{fig:m31}) performed by Garcia et al. (2000, 2001).
The image shows what used to be considered as emission associated with the black hole at the center of the galaxy now resolved into five distinct objects.
A most interesting consequence is that the emission from the region surrounding the central black hole is unexpectedly faint relative to the mass of the central black hole, as with the Milky Way.

\begin {figure} 
\caption {\label{fig:m31} 
Central region of M31 observed with ACIS-I. 
The circle is 5 arcseconds in radius and illustrates the ROSAT HRI location of the nucleus.
The ROSAT source is resolved into 5 individual sources using Chandra, and the source labeled CXO J004244.2+411609 is within 0.15 arcseconds of the 3$\times10^{7}$ solar mass black hole at the nucleus. 
Just to south of the nucleus is the supersoft source CXO J004244.2+411608 (Garcia et al. 2001). 
A long lived transient source, CXO J004242.0+411609, is also shown.
}
\end {figure}

M81 (NGC 3031) is a Sab spiral at a distance of approximately 3.6 Mpc. The galaxy was observed by Tennant et al. (2001b) with the S3 chip of the ACIS-S instrument on Chandra for 50 ksec (Figure~\ref{fig:m81}). Prior to this observation, the galaxy had been observed with both the Einstein and Rosat observatories.  Nine sources were detected with Einstein by Fabbiano (1988) of which 5 were in the spiral arms. Twenty-six sources were detected with ROSAT by Immler and Wang (2001). The Chandra observation detected the bright nucleus at a (0.2-8.0 keV) luminosity of $4\times10^{40}$ ergs/s. In addition, 96 other sources were detected: 81 with S/N $\geq3.5$; 16 with $3.0\leq S/N \leq 3.5$. Here S/N = 3 is one false detection; 3.5 is 0.1. Based on a canonical spectrum, the (0.2-8.0 keV) luminosity of the sources, including the nucleus, ranges from $3\times10^{36}$ ergs/s to $4\times10^{40}$ ergs/s.

There were 41 sources in the bulge of the galaxy and, excluding the nucleus, these had a total luminosity of $1.6\times10^{39}$ ergs/s. One of these sources had a soft spectrum (T$ \simeq 70$ eV) and an observed luminosity in excess of $2\times10^{38}$ ergs/s, about the Eddington limit for a canonical neutron star. 
In addition, the bulge shows $0.8\times10^{39}$ ergs/s of unresolved emission which follows the starlight and is not completely consistent with the extrapolation of the LogN-LogS curve for the galaxy, indicating diffuse emission. 

There were 56 sources in the disk that were within the ACIS-S3 field of view. Twenty one were within $\pm 400$ pc of the spiral arms. 
These 21 include 4 of 5 of the soft sources and 7 of the 10 brightest sources. 
Five of the 21 sources are near SNR. 
Interestingly, 35 disk sources are not in the spiral arms and are typically fainter than those that are, leading Tennant et al. (2001b) to speculate that these are perhaps high-mass x-ray binaries or black-hole candidates. 
There were no associations with any of the 3 known globular clusters in the S3 viewing field. 

\begin {figure} 
\caption {\label{fig:m81} Chandra observations of M81. Left - x-ray image with contours. Right - x-ray contours on optical image. Courtesy of Doug Swartz}
\end {figure}

The sheer number of X-ray sources detected by Chandra in a typical nearby galaxy makes studies of the global properties of the X-ray source populations possible.
For instance, it has been suggested by Sarazin, Irwin, \& Bregman (2001), and supported in a theoretical framework by Wu et al. (2001), that LogN-LogS distributions can be used as a distance indicator for giant elliptical galaxies.
Wu et al. (2001) further show that LogN-LogS distributions can be used as a probe of recent star-formation and of the dynamical history of spiral galaxies.
This is only a sample of the use of x-ray data to place constraints on galaxy evolution.

The remarkable Seyfert 2 galaxy, Circinus, has been observed by Sambruna et al. (2001), Bauer et al. (2001a), and Smith \& Wilson (2001).  
The spectrum of the nuclear region shows a wealth of emission lines including lines of Ne, Mg, Si, S, Ar, Ca, Fe and a very prominent Fe-K line at 6.4 keV (Sambruna et al. 2001).  The emission appears to be the reprocessed radiation from the obscured central source and originates within 60 pc of the object. 
In addition to the very detailed spectrum of the nuclear region, sixteen point sources were detected (Bauer et al. 2001, Smith \& Wilson 2001), several of which exhibit emission lines and are very luminous supernova candidates. 
One bright object has a luminosity of $3.4\times 10^{39}$ erg/s and a strong iron emission line at 6.9 keV with an equivalent width of 1.6 keV. 
An eclipsing x-ray binary was discovered with a period of 7.5 hours and a 0.5-10.0 keV luminosity of $3.7 \times 10^{39}$ ergs/s assuming isotropic emission.  
If not local or beamed, such a high luminosity implies that the x-ray emitting object must have a mass substantially greater than that of a neutron star and is thus likely to be a black hole with a mass of more than 25 M$_{\odot}$. 
These ``ultraluminous'' sources appear to be quite common in nearby galaxies (see also Blanton, Sarazin \& Irwin 2001, Sarazin, Irwin \& Bregman 2000, 2001, Angelini, Lowenstein \& Mushotzky 2001 and Fabbiano, Zezas \& Murray 2001). There appear to be too many of such objects to give much credence to the idea that they are all more local (or more distant) than one might think.
King et al. (2001) argue that the ultraluminous sources might be beamed and discuss a link with microquasars. 
 
\subsubsection{Gravitational lenses}
Another unique application of the excellent imaging properties of Chandra is the study of gravitational lenses, where the image separation is usually only about an arcsecond.  
By comparing different images it is possible to measure differential time delays of temporal changes and obtain an estimate of the Hubble Constant. 
More than a dozen lensed systems have thus far been observed, and one, is shown in Figure~\ref{fig:glens}. 
The four lensed images of the quasar are clearly resolved, and a rapid flare was seen in one of two images (Morgan et al. 2001).  
Several lens systems for which the expected delay is only hours are under study. 
The large magnification of some of the lenses allows the study of objects at very high redshifts that would otherwise not be detectable and the most distant (z=1.4) x-ray jet has recently been detected in one of the images of Q0957+561 (Chartas et al. 2001).

\begin {figure} [htb]
\caption {\label{fig:glens} 
A composite of the deconvolved X-ray image of the gravitational lens RX~J0911.4+0551 (top panel) and the light-curves of the lensed images
A2 (left panel) and A1 (right panel). 
}
\end {figure}

\subsubsection {Clusters of Galaxies} \label{sss:gcimages}

Chandra observations frequently exhibit structures with characteristic angular scales of a few arc seconds in clusters of galaxies which previously were believed to be simple systems.  
Two of the more important types of results involve investigations of the interactions between radio sources and the hot cluster gas in some clusters, and the existence and implications of cold fronts in others.

The Hydra A radio galaxy (3C 218) at a redshift of z = 0.052 is associated with
a relatively poor cluster of galaxies. 
This cluster was observed during the orbital verification and activation phase of the observatory; these very early images contained large areas of low X-ray surface brightness, indicating low density regions or cavities in the intracluster gas.
These cavities and other aspects of the X-ray emission now have been
studied by a number of authors McNamara et al. (2000), David et al. (2001), Nulsen et al. (2001). 
Similar but more dramatic cavities are found in the Perseus Cluster (Fabian et al. 2000); this image is shown in Figure~\ref{fig:gc-perseus}.

\begin {figure} 
\caption {\label{fig:gc-perseus} 
Adaptively smoothed 0.5-7.0 keV Chandra image of the
X-ray core of the Perseus cluster (Fabian et al. 2000).}
\end {figure}

Briefly, the Hydra A cavities are found to coincide with the emission lobes
of the radio source. 
The overall temperature of the cluster gas increases from $\sim$3 keV in the central 10~kpc to $\sim$4 keV at a radius of 200~kpc and then gradually decreases to $\sim$3 keV towards the radial limit of the Chandra observation at about 300~kpc. 
However, the cavities are surrounded by bright rims of enhanced x-ray emission
which are cooler than the cluster gas away from the cavities at comparable radii; this shows that the cavities are not created by expanding radio lobes which shock the surrounding medium, as expected in some earlier models of these sources (e.g. Heinz, Reynolds \& Begelman 1998).

 The cavities probably are in local pressure equilibrium with their surroundings. 
The energy input required to maintain the cavities is comparable to the current
radio emission, and also to that required to substantially inhibit a cooling flow in the inner part of the cluster.

The Hydra A cluster contains an unresolved x-ray source coincident with the radio core. 
The point source x-ray spectra are highly absorbed, which indicates that the  source must be contained within the high column density material found towards the nucleus of the galaxy in VLBI radio observations by Taylor (1996). This limits the size of the X-ray emitting region to $\le$24~pc, (McNamara et al. 2000).

The angular resolution also enables more quantitative studies of the cluster merger process. 
The early images of the cluster A2142 (Markevitch et al. 2000) show two sharp,
bow-shaped shocklike surface brightness features; the surface brightness
is discontinuous on a scale smaller than 5-10~arcseconds. 
However, a detailed investigation shows that the pressure is continuous across the boundary, and that the temperature transition has the opposite sign to that
expected from a shock. 
The most likely explanation is that these edges delineate the dense subcluster cores which have survived a merger and the associated ram pressure stripping by the surrounding gas. 

The image of the cluster A3667 (Vikhlimin, Markevitch, \& Murray 2001a, 2001b) contains a well defined subcluster. 
The pressure jump across the boundary of the subcluster is approximately a factor of two, which indicates that the subcluster relative velocity is about equal to the sound speed of the surrounding medium.  
The transition width in this case is 3.5~arcseconds or less, smaller than the Coulomb mean free path of electrons and protons on either side of the front. 
A model of the magnetic field (Vikhlimin, Markevitch, \& Murray 2001b) needed to suppress the particle diffusion requires a field strength of order 10~$\mu$G. 
These and similar studies enabled by the good angular resolution of the  Observatory should lead to significant improvements in our understanding of cluster merger phenomena.

\subsubsection{The X-Ray Background - The Chandra Deep Surveys}

The first sounding rocket flight that detected the brightest X-ray source in the sky, other than the Sun, also detected a general background of x-radiation (Giacconi et al. 1962). 
The nature of the background radiation has been a puzzle for nearly 40 years, although the lack of distortion of the spectrum of the Cosmic Microwave Background place a strong upper limit to the possibility of a truly diffuse component ($<3$\% , Mather et al. 1990).  
Observations with ROSAT at energies below 2 keV made a major step in resolving a significant fraction (70-80\%) into discrete objects (Hasinger et al. 1998)  and found that the sources reside mainly in AGN at redshifts from 0.1 to 3.5.  
ASCA satellite observations extended the search for sources in the 2-10 keV band, resolving about 30\% into mainly AGNs (Ueda et al. 1998). Observations with Beppo-Sax have continued these studies. 
Currently two 1-Ms exposures have been accomplished with Chandra - the Chandra Deep Fields North (Figure~\ref{fig:cdfn}, Brandt et al. 2001b) and South (Giacconi et al. 2001).  
These surveys extend the study of the background to flux levels more than an order of magnitude fainter in the 0.5-2.0 keV band and resolve over 90\% of the background into a variety of discrete sources. 
The largest uncertainty in establishing the fraction is now in the knowledge of the total level of the background itself.   

\begin {figure}
\caption {\label{fig:cdfn} 
Chandra "true -color" ACIS image of the Chandra Deep Field - North.  
This image has been constructed from the 0.5 -2.0 keV band (red) and 2.0 - 8.0 (blue) images (Brandt et al. 2001b).
The location of the HST deep field is shown in outline.
Two of the red diffuse patches may be associated with galaxy groups (Bauer et al. 2001b).
}
\end {figure}

The spectrum of the X-ray background has been called a ``spectral paradox'' by  Boldt (1987), because the majority of the bright AGN are found to have a photon index larger than that for the background itself (1.7 vs 1.4).   
One of the purposes of the Chandra Deep Surveys has been to explore the spectra of the sources.   
Since it must be the faint sources that modify the spectrum from the average  of the bright AGN sample, the faint source spectrum was estimated by adding the spectra of individual sources detected in the surveys.  
Garmire et al. (2001) found the spectral index to decrease from 1.8 to 1.0 as the flux of the sources decreased from $1\times10^{-14}$ to $2\times 10^{-15}$ ergs/cm$^{2}$/s. 
Similar results are found in the data from the southern field. 
The spectrum of all of the sources has a slope very near to 1.4 (Tozzi et al. 2001, Garmire et al. 2001), consistent with the wide field-of-view measurements made with detectors that could not resolve the fainter sources.

In each of the Chandra Surveys about 350 sources were detected.
The flux levels attained in the soft and hard bands were ($3\times10^{-17}$ ergs/cm$^{2}$/s and $2\times10^{-16} $ ergs/cm$^{2}$/s respectively.      
The highest redshift detected (so far) is a QSO at 5.2. 
A highly obscured type 2 QSO at z = 3.4 has been reported in the Chandra Deep Field - South by Norman et al. (2001). 

Since these fields were selected not to have any nearby galaxies in them, the nearest galaxies detected are at redshifts of $\sim 0.08$.  
In one of these galaxies, located in the HDF-N, the x-ray emission appears to be non-nuclear, based on the offset from the center of the optical light distribution, perhaps implying on the basis of the high x-ray luminosity that intermediate mass black hole candidates are present Hornschemeier et al. (2000). 

The LogN - LogS function has a change in slope at just over $1\times10^{-14}$ ergs/cm$^{2}$/s for the 2-10 keV band and at $\sim 5\times10^{-15}$ ergs/cm$^{2}$/s for the 0.5-2.0 keV band. 
A nominal slope of -1.5 would be expected for a Euclidean geometry populated with a uniform distribution of sources.  
In the hard band, the slope flattens to -1 at about $1.4\times10^{-15}$ ergs/cm$^{2}$/s and then becomes even flatter at lower fluxes (Figure~\ref{fig:lnlshard}). 
In the soft band the slope flattens from -1.5 to -0.65 down to $1\times10^{-16}$ ergs/cm$^{2}$/s, then becomes even flatter for fainter fluxes (see Figure \ref{fig:lnlssoft}).  
Eventually, these curves should change to steeper slopes once the population of galaxies is reached,  
Normal galaxies will not contribute very much to the total X-ray background flux, since they are so faint but will ultimately become the major contributor at the lowest flux levels (Hornschemeier et al. 2002).  

\begin {figure}
\caption {\label{fig:lnlshard} 
The integral LogN-LogS plot for the Chandra Deep Field North of the hard-band sources (small solid squares) found in in three areas of the field.
The portion of the plot above $10^{-14}$ ergs/cm$^{2}$ s deg$^{2}$ was taken
from the full image, the portion between $10^{-15}$ and $10^{-14}$ ergs/cm$^{2}$ s deg$^{2}$ from a 6 arcmin radius region, and the faintest sources from a region of 3 arcmin radius.  
The large open square and the dash-dot "bow tie" are from Ueda et al. (1998) and Gendreau (1998); the small open squares are also from the ASCA Large Sky Survey (Ueda et al. 1998), and the solid circle and dotted "bow tie" region are from the Ginga survey and fluctuation analysis respectively (Hayashida, Inoue, \& Kii 1991).  
}
\end {figure}

\begin {figure}
\caption {\label{fig:lnlssoft} 
The integral LogN-LogS plot for the Chandra Deep Field-North of the soft-band sources found in the same three areas as for the hard-band data.    
The open circles and the dashed region are from ROSAT (Hasinger et al. 1998). 
}
\end {figure}

The majority of the x-ray sources beyond a redshift of 0.5 are AGN or QSOs.  
Barger et al. (2001) found, based on the luminosities of the hard-band X-ray sources, that the accretion rate onto black holes grows linearly with redshift to a redshift of 1 and then flattens with only a slight increase out to a redshift of 3. 
The volume density of black hole accretion is found to increase as (1+z)$^{3}$.

\subsubsection{Gamma-Ray Bursts}
The moderately rapid response for targets of opportunity has made possible the study of the afterglows of gamma-ray bursts with the Observatory.  
The afterglow of GRB991216 showed the first X-ray iron line profile indicating a very high velocity ($\sim 0.1$c) in the ejected material (Piro et al. 2000). 
These authors also reported a recombination edge from hydrogenic ions of iron at a redshift of 1. 
This observation supports a hypernova interpretation, or delayed gamma-ray burst following a supernova (Vietri and Stella 1998, Meszaros and Rees 2001).  A second gamma-ray burst, GRB000926, revealed an unusual x-ray light curve that implied that the burst expanded into a dense medium (n$\sim 10^{4}$ cm$^{-3}$) and that the fireball was only moderately collimated initially which then slowed down and become non-relativistic after 5 days (Piro et al. 2001). 

\subsection {High Resolution Spectroscopy} \label{sss:scispec}

Owing to their unprecedented clarity, Chandra images are visually striking and provide new insights into the nature of x-ray sources.
Equally important are Chandra's unique contributions to high-resolution dispersive spectroscopy.

High-resolution x-ray spectroscopy is the essential tool for diagnosing conditions in hot plasmas.
It provides information for determining the temperature, density, elemental abundance, and ionization stage of x-ray emitting plasma.
The high spectral resolution of the Chandra gratings isolates individual spectral lines which would overlap at lower resolution.
The high spectral resolution also enables the determination of flow and turbulent velocities, through measurement of Doppler shifts and widths.
Dispersive spectroscopy achieves its highest resolution for spatially unresolved (point) sources.
Thus, Chandra grating observations have concentrated on, but are not limited to, stellar coronae, x-ray binaries, and active galactic nuclei.

\subsubsection {Stellar Coronae} \label{ssss:starspec}

The spectra of stellar coronae obtained with Chandra contain a large number of interesting emission line features that serve as diagnostic tools for temperatures, densities, and emission measures. 
Figure~\ref{fig:capellaandprocyon} shows the LETGS spectra (5-175 \AA) of Capella and Procyon, two different coronal sources. 
There are a large number of lines from very many different elements, and also a strong temperature dependence. 
In Capella there are many lines around 15 \AA\ from Fe XVII, while these lines are weak in the Procyon spectrum.
Conversely, the Fe IX line at 171 \AA, is very prominent in the Procyon spectrum indicating a cooler corona. 

\begin {figure} 
\caption {\label{fig:capellaandprocyon} 
The LETGS spectra of the two stars Capella (left) and Procyon (right).}
\end {figure}

Apart from a chemical and temperature analysis, one can derive densities from many density-sensitive lines. 
In the wavelength range around 100~\AA\ of the Capella spectrum, density-sensitive lines of highly ionized Fe-ions appear and in the short-wavelength region (between 6 and 45 \AA) of both spectra, temperature- and density-sensitive lines are present. 
The latter originate from the He-like ions Si XIII, Mg XI, Ne IX, O VII, N VI, and C V.
For these ions the resonance line (r) 1s$^{2\ 1}$S$_0$ -- 1s2p $^1$P$_1$,
the intercombination line (i) 1s$^{2\ 1}$S$_0$ -- 1s2p $^3$P$_{1,2}$ and the forbidden line (f) 1s$^{2\ 1}$S$_0$ -- 1s2s $^3$S$_1$ are resolved. 
 
From the resonance lines one can obtain an estimate of the temperature --- an estimate as the flux may arise from two different regions on the stellar surface.  
In the He-like ions, the ratio between the intercombination line and the forbidden line is strongly density dependent, while the ratio between the the sum of the intercombination line and forbidden line and the resonance line is temperature dependent.
In a low-density plasma the forbidden line is stronger than the intercombination line, but for increasing density, the 1s2s $^3$S$_1$ -- the upper level of the forbidden transition -- will be depopulated by collisions in favour of the 1s2p $^3$P$_{1,2}$ -- the upper level of the intercombination line.
The resonance line intensity is comparable to the sum of the intensities
of the two other lines and increases at higher temperatures.
Table~\ref{table:temp} shows the temperature for the coronae of Capella and Procyon, based on the ratio of the sum of the intercombination and forbidden lines to the resonance line and also on the ratio between lines of succeeding stages of ionization. 
From this table we notice that the coronae of both stars have a multi-temperature structure and that the temperature of the corona of Capella is higher and extends to higher ionization stages.
Table ~\ref{table:density} shows the densities derived from the appropriate line ratios of the He-like ions.

\begin{table}
\caption{Temperature determination from the line intensity ratio $(i+f)/r$ for different He-like ions and from the ratio of the He- and H-like resonance lines. 
Temperatures (in MK) derived from SPEX/MEKAL and Porquet et al. (2001).
For Capella. the data are from Mewe, Kaastra, \& Drake (2001), and for Procyon from Raassen et al. (2001).}                                              
\begin {center}
\label{table:temp}
\begin{tabular}{l|cc}
Ion         & Capella    & Procyon\\
\hline
$C V$ & 1.4$\pm{0.5}$& 0.4  \\
$N VI$ & 0.5$^{+0.5}_{-0.2}$&1.2 \\
$O VII$ & 1.8$\pm0.3$&1.1  \\ 
$Mg XI$ & 4.6$^{+1.4}_{-1.0}$& -\\ 
$Si XIII$ & 5$^{+3}_{-2}$&- \\
\hline 
$C V/C VI$ &1.10$\pm0.15$& 1.12  \\ 
$N VI/VII$ &2.50$\pm0.14$& 1.62 \\ 
$O VII/VIII$ &3.37$\pm0.06$& 2.14 \\ 
$Mg XI/XII$ &6.9$\pm0.2$& - \\ 
$Si XIII/XIV$ &9.4$\pm0.5 $& - \\
$Fe X/IX$ & - &1.25 \\
\end{tabular}
\end{center}
\end{table}

\begin{table}
\caption{Density determination from the $i/f$ line intensity ratio for different He-like ions.
Electron density (in cm$^{-3}$) derived from SPEX/MEKAL and Porquet et al. (2001). The Capella data are from  Mewe, Kaastra, \& Drake (2001); 
the Procyon data are from Raassen et al. (2001).
}                                              
\begin {center}
\label{table:density}
\begin{tabular}{l|cc}
Ion         & Capella    & Procyon\\
\hline
$C V$ & 3$\pm 2$~$10^9$ & 4\ 10$^9$    \\
$N VI$ & 5$\pm 3$~$10^9$ & 1\ 10$^{10}$\\
$O VII$ & $<7\ 10^{9}$& $6\ 10^{9}$\\
$Mg XI$   &  $<4\ 10^{12}$ & - \\
$Si XIII$   & 3$\pm 2.5$~$10^{13}$&  -  \\
\end{tabular}
\end{center}
\end{table}

\subsubsection {X-ray Binaries} \label{sss:xrbspec}

The x-ray output of the bright Galactic x-ray binaries is generally dominated
by continuum from an optically thick accretion disk, but grating spectra
are revealing a rich variety of absorption and emission features that
carry new information about material in and around the source (e.g., 
Brandt \& Schulz (2000), Paerels et al. (2001), Cottam et al. (2001), Marshall, Canizares, \& Schulz (2001), and Schulz (2001).  
Doppler structure is often seen in the emission and/or absorption 
lines, with velocities ranging from hundreds of km/s (e.g. for 4U 1822-37 where the lines are attributed to recombination in an x-ray illuminated bulge where
the accretion stream hits the disk; (Cottam et al. 2001) up to 0.26c for emission lines in relativistic jets of the binary SS433 (Marshall et al. 2001).  
One interesting example is the mysterious binary Circinus X-1, thought to contain a neutron star, that at times radiates beyond its Eddington limit (Brandt \& Schulz, 2000).  
The HETG spectra reveal lines from H-like and/or He-like Ne, Mg, Si, S and Fe. 
The lines exhibit broad ($\pm$2000 km/s) P Cygni profiles, with blue-shifted
absorption flanking red-shifted emission.  
Examples are shown in Figure~\ref{fig:cirx1fig3}. 
Brandt \& Schulz (2000) interpret these features as the X-ray signatures of a wind being driven off the accretion disk, making Cir X-1 the X-ray analog of a broad absorption line quasar. 

Line ratios are being used to constrain physical properties of the emitting regions, and time variability of intensities and Doppler structures helps to
locate the source of the lines in the binary system. 
The HETG spectrum of the ultra-compact binary 4U 1626-67 shows
unexpectedly strong photoelectric absorption edges of Ne and O, most
probably from cool, metal rich material local to the source  (Schulz et al. 2001). 
The anomalous abundances led the authors to suggest
that the mass donor in this binary is the chemically fractionated core of a C-O-Ne or O-Ne-Mg white dwarf.

\begin{figure}
\caption{\label{fig:cirx1fig3}
Several of the strongest X-ray P Cygni profiles seen from Cir X-1
with the HETG (Brandt \& Schulz 2000). The two middle panels show independent
spectra of the Si {\sc XIV} profile with the HEG ($1^{\sc st}$ order)
and MEG ($3^{\sc rd}$ order). Typical bins have 200-1200 counts. The
velocity in each panel gives the instrumental resolution; the lines
are clearly broader}
\end{figure}

\subsubsection {Active Galactic Nuclei (AGN)} \label{ssss:agn}

High resolution spectra of AGN, especially Seyfert Galaxies, are providing extraordinary new details about the physical and dynamical properties of material surrounding the active nucleus. 
In the case of Seyfert 1 galaxies, whose signal is dominated by the bright X-ray continuum from the central engine, the partially ionized circum-source material introduces a prominent pattern of absorption lines and edges. 
These are the "warm absorbers" originally discovered in low resolution spectra by Reynolds (1997), and George et al. (1998) but now revealed in much greater detail.  

For example, the LETGS spectrum of NGC~5548 shown in (Figure~\ref{fig:5548letg}) exhibits three dozen absorption lines, plus a few in emission (Kaastra et al. 2000), and the HETGS spectrum of NGC~3783 (Kaspi et al. 2000, 2001) has over five dozen lines.
In both cases, there is evidence for bulk motions of several hundred km s$^{-1}$. 
For NGC~3783, detailed modeling of an absorber with two ionization
components does a remarkably good job of reproducing the observed line
strengths (Kaspi et al. 2001). 
The HETGS spectrum of MCG-6-30-15  (Figure~\ref{fig:4151_mcg6hetg}) also has
dozens of absorption lines from a wide range of ionization states (Lee et al. 2001).
In addition, an absorption feature at ~0.704 keV (17.61 \AA) is well fit by the neutral Fe L3 absorption edge (and associated resonant structure) with a column density equal to the amount of line-of-sight dust deduced from earlier reddening studies. (Alternatively, Branduardi-Raymond et al. (2001) attribute this feature to relativisticly broadened
O{\sc VIII} emission, based on their XMM-Newton RGS spectrum).
The HETG also has been used to (marginally) resolve a narrow component of the Fe K$\alpha$ line in NGC~5548 (Yaqoob 2001).

\begin {figure} 
\caption {\label{fig:5548letg}
LETGS spectrum of the Seyfert 1 galaxy NGC~5548, corrected for order contamination, redshift and Galactic absorption (Kaastra et al. 2000). Several prominent absorption lines from H-like and He-like ions are marked, as is
the forbidden line of He-like oxygen. 
}
\end {figure}

\begin{figure}
\caption {\label{fig:4151_mcg6hetg} 
A portion of the HETG spectra of the Seyfert 1 galaxy MCG-6-30-15: top, corrected for instrumental effective area; see Lee et al. (2001) and 
the Seyfert 2 galaxy NGC~4151 (bottom, in counts per bin; see 
(Ogle et al. 2001)). 
The top spectrum also shows the best fit power law including galactic absorption but excluding the region most affected by the warm absorber) and the magnitude of the one-sigma uncertainty.}
\end{figure}

For Seyfert 2's the strong continuum from the central engine is not seen directly, so the surrounding photo-ionized regions are seen in emission. 
The HETGS spectra of Markarian 3 (Sako et al. 2000) and NGC~4151 (classified as
Sy 1.5 but observed when the continuum was exceptionally faint, (Ogle et al. 2001) bristle with emission lines, whose ratios provide diagnostics of the conditions in the emitting clouds (Figure~\ref{fig:4151_mcg6hetg}).  
There are clear signatures of photoionization, such as the relatively strong forbidden line from He-like ions and narrow features from free-bound radiative recombination.
Other diagnostics suggested the presence of a smaller amount of collisionally excited plasma (Ogle et al. 2001), although more recent observations and modeling
indicate that photoexcitation together with the photoionization and  recombination can explain the line ratios very well (Sako et al. 2000).

Many AGN spectra show primarily a strong, power-law continuum with little
or no evidence for any absorption or emission lines. 
Examples include the narrow-line Seyfert 1 galaxy Ton~S180  (Turner et al. 2001), BL Lac objects and radio-loud quasars (Fang et al. 2001).

\subsubsection {Young Supernova Remnants} \label{ssss:ysnr}

Since the X-ray output of a young SNR is dominated by a moderate number of
strong emission lines, the dispersed spectrum resembles a spectroheliogram,
showing multiple images of the remnant in the light of individual lines.
Such is the case for 1E0102-72, a SNR in the SMC estimated to be $\sim 1000$
years old (Figure \ref{fig:e0102hetg}: Canizares et al. 2001; Flanagan et al.2001).
The monochromatic images are dominated by emission from the shocked
stellar ejecta. 
For a given element, the images for the He-like resonance
are systematically smaller than those for the H-like Ly$\alpha$ line, which
is graphic evidence for the progression of the so-called reverse shock backwards into the expanding ejecta (also seen in the ACIS image by Gaetz et al. 2000).
Doppler shifts of $\sim \pm 2000$ km s$^{-1}$ are measured by comparing the 
strongest monochromatic images from the plus and minus orders.
The velocities appear asymmetric, suggesting that the shock
heated ejecta fill a torroidal region inclined to the line of sight. 

\begin{figure}
\caption{\label{fig:e0102hetg}
A portion of the dispersed HETG spectrum of the SNR E0102-72.
}
\end{figure}

\section {Conclusion} \label{s:conc}

The Chandra X-Ray Observatory is performing as well, if not better, than anticipated~--- and the results are serving to usher in a new age of astronomical and astrophysical discoveries.

\section {Acknowledgements} \label{s:ackn}

We recognize the efforts of the various Chandra teams which have contributed so much to the success of the observatory.
In preparing this overview, we have used figures and material drawn from their work. We would especially like to thank Dr. Douglas Swartz and Professor J. Grindlay for their contributions and assistance. 

\appendix
\section{Chandra web sites} \label{app:websites}
The following lists several Chandra-related sites on the World-Wide Web (WWW).

\begin{description}
\item[http://chandra.harvard.edu/] Chandra X-Ray Center (CXC), operated for NASA by the Smithsonian Astrophysical Observatory.
\item[http://wwwastro.msfc.nasa.gov/xray/axafps.html] Chandra Project Science, at the NASA Marshall Space Flight Center.
\item[http://hea-www.harvard.edu/HRC/] Chandra High-Resolution Camera (HRC) team, at the Smithsonian Astrophysical Observatory (SAO).
\item[http://www.astro.psu.edu/xray/axaf/axaf.html] Advanced CCD Imaging Spectrometer (ACIS) team at the Pennsylvania State University (PSU).
\item[http://acis.mit.edu/] Advanced CCD Imaging Spectrometer (ACIS) team at the Massachusetts Institute of Technology.
\item[http://www.sron.nl/missions/Chandra]  Chandra Low-Energy Transmission Grating (LETG) team at the Space Research Institute of the Netherlands.
\item[http://www.ROSAT.mpe-garching.mpg.de/axaf/] Chandra Low-Energy Transmission Grating (LETG) team at the Max-Planck Instit\"ut f\"ur extraterrestrische Physik (MPE).
\item[http://space.mit.edu/HETG/] Chandra High-Energy Transmission Grating (HETG) team, at the Massachusetts Institute of Technology.
\item[http://hea-www.harvard.edu/MST/] Chandra Mission Support Team (MST), at the Smithsonian Astrophysical Observatory.
\item[http://ipa.harvard.edu/] Chandra Operations Control Center, operated for NASA by the Smithsonian Astrophysical Observatory.
\item[http://ifkki.kernphysik.uni-kiel.de/soho] EPHIN particle detector.
\end{description}

\clearpage

\end{document}